\documentclass[rnote]{aa} 

\usepackage{graphicx}
\usepackage{txfonts}
\begin{document}

   \title{HCN observations of comets C/2013 R1 (Lovejoy) and C/2014 Q2 (Lovejoy)} 

   \author{E.S. Wirstr\"{o}m\inst{1} \and
          M.S. Lerner\inst{1} \and
          P. K\"{a}llstr\"{o}m\inst{2} \and
          A. Levinsson\inst{3} \and
          A. Olivefors\inst{2} \and
          E. Tegehall\inst{2} 
          }

   \institute{Department of Earth and Space Sciences, Chalmers University of Technology, Onsala Space Observatory, SE-439 92 Onsala, Sweden,
              \email{eva.wirstrom@chalmers.se}
             \and
             Department of Chemistry and Chemical Engineering, Chalmers University of Technology, SE-412 96 Gothenburg, Sweden 
             \and
             Department of Applied Physics, Chalmers University of Technology, SE-412 96 Gothenburg, Sweden
             }

   \date{Received 30 September, 2015; accepted 19 February, 2016}

  \abstract{HCN J=1--0 emission from the long-period comet C/2013 R1 (Lovejoy) was observed from the Onsala Space Observatory on multiple occasions during the month before its perihelion passage on December 22, 2013. We report detections for seven different dates, spanning heliocentric distances ($R_h$) decreasing from 0.94 to 0.82 au. Estimated HCN production rates are generally higher than previously reported for the same time period, but the implied increase in production rate with heliocentric distance, $Q_{\rm HCN}  \propto R_h^{-3.2}$, represent well the overall documented increase since it was first observed at $R_h$=1.35. The implied mean HCN abundance relative to water in R1 Lovejoy is 0.2\%.   
We also report on a detection of HCN with the new 3 mm receiver system at Onsala Space Observatory in comet C/2014 Q2 (Lovejoy) on January 14, 2015, when its heliocentric distance was 1.3 au. Relative to comet C/2013 R1 (Lovejoy), the HCN production rate of C/2014 Q2 (Lovejoy) was more than 5 times higher at similar heliocentric distances, and the implied HCN abundance relative to water 0.09\%.}


   \maketitle
%

\section{Introduction}
Comets, bodies of molecular ices and crystalline silicate dust, were formed far out from the Sun in the young solar system, and the vast majority have remained there ever since. Their compositions can therefore provide clues to the local conditions during the formation of the Solar System \citep{MummaCharnley11,BockeleeMorvan04}. When a cometary orbit is perturbed, bringing it within about 3 au from the Sun, the ices in its nucleus sublimate to form a coma of gaseous volatiles and dust, allowing for remote investigations of their composition. Spectroscopic investigations at infrared \citep[e.g. ][]{DiSanti13} and radio wavelengths \citep[e.g. ][]{Remijan08,Biver12} have revealed a wide chemical diversity among comets. Thus, detailed studies of cometary comae are essential to explore the extent of heterogeneity amongst comets, and provide further understanding of the distribution of molecular material throughout the Solar System during its formation.

The water production rate of a comet is expected to scale approximately as $R_h^{-2.0}$, where $R_h$ is the comet's radial distance to the Sun.
HCN is one of the cometary molecules whose production rate follows most closely that of water, typically with a mixing ratio close to 0.1\%, and is therefore commonly used as a proxy to the water production rate \citep{MummaCharnley11}.
By combining observations of HCN over a range of heliocentric distances, the production rate variability can be studied and possibly connected to outgassing mechanisms, physical structure, and distribution of volatiles near the nucleus surface.

We report on observations of hydrogen cyanide (HCN) in comet C/2013 R1 (Lovejoy) (hereafter referred to as R1 Lovejoy) and C/2014 Q2 (Lovejoy) (hereafter Q2 Lovejoy). Comet R1 Lovejoy is a high-inclination ($i$=64$\degr$), long-period comet with orbital period of about 7000 years. It was discovered by Terry Lovejoy on September 7, 2013 at a heliocentric distance of 1.98 au \citep{Lovejoy13}, and passed perihelion at 0.81 au on December 22, 2013. Comet Q2 Lovejoy was also discovered by Terry Lovejoy, but almost a year later, on August 17, 2014 \citep{Lovejoy14}, at a heliocentric distance of 2.63 au. It is also a long-period, high-inclination ($i\sim$80$\degr$) comet which reached perihelion on January 30, 2015, at a closest heliocentric distance of 1.29 au.


\section{Observations and data reduction}
The $J$\,=\,1\,--\,0 transition of HCN at 88.63 GHz was observed towards comet C/2013 R1 (Lovejoy) using the 3~mm receiver at the Onsala Space Observatory 20 meter antenna (OSO20m) on multiple occasions during November and December 2013. At the time of observations, comet R1 Lovejoy was approaching perihelion, but receding from Earth. The geometric circumstances for the observing dates are included in Table~\ref{ObsTable}.

A frequency switching mode with a switching frequency of 5 Hz and a throw of 5 MHz was used, and spectra were recorded with a 1600 channel correlator at channel separation 12.5~kHz, thus providing 20 MHz bandwidth. Before each observing session, tabulated hourly ephemerides from the latest available solution were obtained from the Jet Propulsion Laboratory Horizons system\footnote{\texttt{http://ssd.jpl.nasa.gov/horizons.cgi}}, which were then interpolated by the telescope control system. The pointing accuracy was typically checked after sunrise and sunset, and its magnitude was not found to vary significantly between observing days. Pointing offsets of less than 5\arcsec\ were found and applied to the pointing model, which has an rms of 3\arcsec. The system temperature varied from 300~K and up during and between the observing days -- spectra with system temperatures above 1000~K have been excluded from the data set. The total integration times for spectra included in the analysis for each day are given in Table~\ref{ObsTable}.  Baselines of polynomial order 3, 5 or 7 were fitted to individual spectra and subtracted before averaging spectra using the recorded system temperatures as weights. 

The same transition of HCN was observed towards comet C/2014 Q2 (Lovejoy) on January 14, 2015, but with the new dual-polarisation, sideband-separating 3-mm receiver system\footnote{\texttt{http://www.chalmers.se/en/centres/oso/radio\-astronomy/20m/Pages/Description.aspx}} \citep{Belitsky15} in dual beam-switching mode (DBSW) with secondary beam offset 11\arcmin. The system temperature varied between 140--400~K and spectra were recorded with a Fast Fourier Transform spectrometer at channel separation 12.2 kHz, covering 100 MHz bandwidth. The use of DBSW together with the fact that the new system is more stable, made it sufficient to fit and subtract linear baselines from individual spectra before averaging with the same method as for R1 Lovejoy.

At 88.6 GHz the OSO20m beam Full Width at Half Maximum (FWHM) is about 44\arcsec\, corresponding to about 1.3 -- $2.5\times 10^4$~km at the distance of comet R1 Lovejoy and $1.6\times 10^4$~km at the distance of Q2 Lovejoy during observations. The main beam efficiency is about 0.53 at this frequency. Before analysis and presentation, reduced spectra were redressed to a channel spacing of 0.2~km\,s$^{-1}$.

\begin{table*}
\caption{Summary of observations and results}             
\label{ObsTable}      
\centering          
\begin{tabular}{c c c c c c c c c c} 
\hline\hline       

Date & $R_h$ & $\Delta$ & $t_{int}$ & $\int T_{\rm mb}$ d$v$ & $T_{\rm K}$ & $Q_{\rm{H}_2\rm{O}}$ & $N_{\rm HCN}$ & $Q_{\rm HCN}$ \\ 
 & (au) & (au) & (h) & (mK km\,s$^{-1}$) & (K) & (10$^{29}$ s$^{-1}$) & (10$^{11}$ cm$^{-2}$) & (10$^{26}$ s$^{-1}$) \\
 \hline    
2013 & \multicolumn{8}{c}{R1 Lovejoy} \\ 
\hline               
22 Nov   & 0.986 &    0.404 & \phantom{0}4.4 & 213$\pm$32 & 62\tablefootmark{a} & 0.7\tablefootmark{b} & 8.38$\pm$1.26 & 1.9$\pm$0.3 \\ [3pt]
23 Nov  & 0.980 &    0.409 & 12.1 & \phantom{0}74$\pm$15 & 62\tablefootmark{a} & 0.7\tablefootmark{b} & 2.92$\pm$0.58 & 0.7$\pm$0.1 \\ [3pt]
24 Nov   & 0.971 &    0.415 & 11.3 & 158$\pm$13 & 63\tablefootmark{a} & 0.7\tablefootmark{b} & 6.21$\pm$0.52 & 1.5$\pm$0.1  \\ [3pt]
30 Nov   & 0.915 &    0.483 & \phantom{0}7.3 & 149$\pm$24 & 66\tablefootmark{a} & 0.8\tablefootmark{b} & 5.93$\pm$0.96 & 1.7$\pm$0.3  \\ [3pt]
\phantom{0}1 Dec   & 0.906 &    0.499 & \phantom{0}7.4 & 196$\pm$21 & 67\tablefootmark{a} & 0.8\tablefootmark{b} & 7.78$\pm$0.82 & 2.4$\pm$0.3 \\ [3pt]
\phantom{0}2 Dec  & 0.898 &    0.514 & 10.4 & 111$\pm$18 & 69\tablefootmark{a} & 0.9\tablefootmark{b} & 4.50$\pm$0.74 & 1.4$\pm$0.2 \\ [3pt]
15 Dec   & 0.823 &    0.763 & \phantom{0}6.8 & 100$\pm$28 & 81\tablefootmark{a} & 1.1\tablefootmark{b} & 3.97$\pm$1.12 & 2.2$\pm$0.6 \\ [3pt]
\hline       
2015 & \multicolumn{8}{c}{Q2 Lovejoy} \\         
\hline
14 Jan & 1.312 & 0.503 & \phantom{0}6.3 & 432$\pm$13 & 73 & 5 & 21.3$\pm$0.7 & 4.5$\pm$0.1 \\ [3pt] %
\hline 
\end{tabular}
\tablefoot{
\tablefoottext{a}{From extrapolation of $T_{\rm K}$ measurements by \citet{Biver14}, see text for details.}
\tablefoottext{b}{From extrapolation of $Q_{\rm{H}_2\rm{O}}$ values from \citet{Biver14}.} 
}
\end{table*}

\section{Results and Analysis}
\subsection{R1 Lovejoy}
\begin{figure}
   \centering
   \includegraphics[width=\hsize]{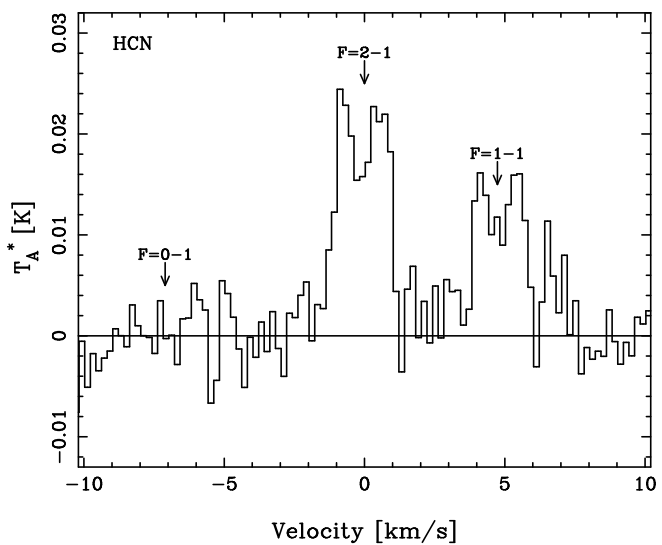}
   \caption{HCN total average spectrum for C/2013 R1 (Lovejoy), including data from all detection days. The three hyperfine components are marked with arrows.}
              \label{FigSpectrum}
                  \end{figure}

The main beam intensities, integrated over all three HCN hyperfine components and with associated 1$\sigma$ errors, are given in Table~\ref{ObsTable} for each observing day, and Figure~\ref{FigSpectrum} shows the total average R1 Lovejoy HCN spectrum including data from all the observing dates. The two strongest hyperfine components show clear double peaks. However, the relative intensities are fairly similar so no conclusions can be drawn about outgassing asymmetries.  

Based on observations of a set of methanol rotational transitions around 251~GHz, \citet{Biver14} report on the coma kinetic temperatures for R1 Lovejoy on November 8 (55~K), November 27 (65~K) and December 9 (80~K), 2013. Since no temperature information can be derived from the current dataset, these data from concurrent observations at a similar frequency, thus probing the same part of the coma, should provide appropriate estimates of the excitation conditions for HCN. As the relation between the three reported temperatures are not at all well described by a power law, the coma kinetic temperatures on the dates of observations are approximated by linear interpolations between adjacent data points, see Table~\ref{ObsTable}. The water production rates, $Q$(H$_2$O), presented in Table~\ref{ObsTable} are derived in the same way, from the published water production rates in \citet{Biver14}.  
Spectroscopic parameters for HCN in its vibrational ground-state are adopted from the Cologne Database for Molecular Spectroscopy\footnote{\texttt{http://www.ph1.uni-koeln.de/vorhersagen/}} \citep{Muller05}, and the HCN 1--0 emission is assumed to be optically thin, as verified by the ratio of line intensities. 

\subsubsection{Column densities and production rates}
The rotational excitation of molecules is dominated by collisions in the inner parts of the coma, but with increasing distance from the nucleus densities decline and radiative processes become more important. For HCN, the excitation cannot be assumed to be thermalised beyond radii of $5\times10^3$ km \citep{BockeleeMorvan04}, meaning that local thermodynamical equilibrium (LTE) analysis of the current observations, sampling coma gas out to $\sim10^4$ km, will be inadequate.

Therefore, to estimate the HCN production rate, $Q_{\rm HCN}$, and beam averaged column densities, we have used the excitation and radiative transfer model described in \citet{Biver99}. The model assumes a spherically symmetric Haser model for the density distribution of molecules in the coma, and considers collisional excitation by water and electrons, as well as radiative excitation. It assumes a density distribution of parent molecules in the cometary coma proportional to $Q e^{-r/(v_{\rm exp}t_{\rm l})} / (v_{\rm exp} r^2)$, where $v_{\rm exp}$ is the constant gas expansion velocity and $t_{\rm l}$ is the molecule lifetime. In this model, the excitation goes from being collision-dominated in the inner part of the coma, to being dominated by the balance between solar pumping and spontaneous decay at large enough radii. For more details on e.g. used collision rates and excitation models, see \citet{Biver99} and references therein. 

The gas expansion velocity is expected to increase as the comet approaches the Sun, but no such trend could be securely determined from our dataset. Instead, it was taken as constant and estimated from the line profiles in the total averaged spectrum shown in Fig.~\ref{FigSpectrum}, namely as the average of the half-widths at half-maximum intensity (HWHM) for the two strongest hyperfine line components, giving $v_{\rm exp}=1.1\pm0.1$~km\,s$^{-1}$.  Since HCN lifetimes are quite sensitive to solar activity \citep{BockeleeMorvanCrovisier85}, and the observations were done at a time of moderate solar activity, a photodissociation rate of $2.0\times10^{-5}$~s$^{-1}$ was adopted based on \citet{Crovisier94}, following the example of \citet{Biver00}. This rate was scaled based on the average heliocentric distance for each observing day. The resulting HCN production rates fall between 0.7 and $2.4\times10^{26}$~molecules per second and are listed in the rightmost column of Table~\ref{ObsTable}.

\subsection{Q2 Lovejoy}
\begin{figure}
   \centering
   \includegraphics[width=\hsize]{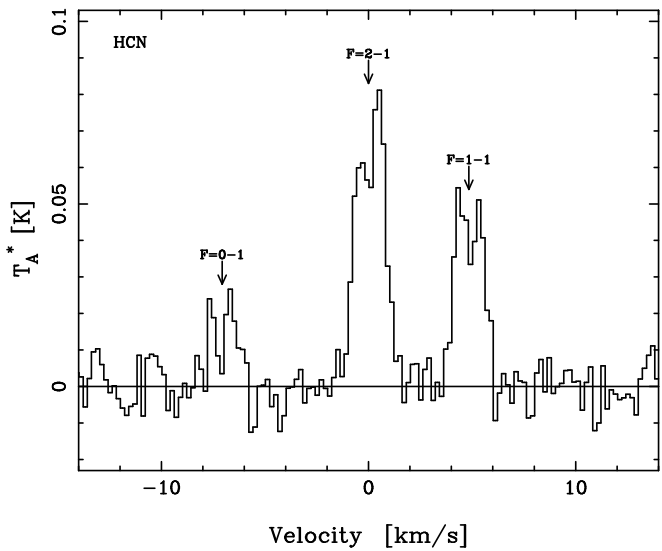}
   \caption{Average HCN spectrum for C/2014 Q2 (Lovejoy) on January 14, 2015. The three hyperfine components are marked with arrows.}
              \label{FigSpectrumQ2}
    \end{figure}

Figure~\ref{FigSpectrumQ2} shows the total averaged HCN spectrum towards Q2 Lovejoy and the resulting integrated intensity is included in Table~\ref{ObsTable}. The signal is stronger than for any of the dates R1 Lovejoy was observed -- all three hyperfine components are clearly detected at the expected line ratios of 1:5:3. Moreover, the measured integrated intensity is more than twice compared to R1 Lovejoy at highest intensity -- although Q2 Lovejoy at the time of observation was further from both the Sun and the Earth. The noise level reached in just above six hours on-source time with the new receiver system is similar to what was obtained from more than 10 hours on-source time for R1 Lovejoy in 2013. 

From observations of a suite of methanol lines between January 13--26 2015, \citet{Biver15} derive a gas kinetic temperature of 73 K for the coma of comet Q2 Lovejoy. The study also provides a water production rate of $5\times 10^{29}$ molecules s$^{-1}$ for January 13--16. Adopting an expansion velocity of 0.82$\pm$0.04~km\,s$^{-1}$ from the line profile, the HCN production rate is then estimated to $4.5\times 10^{26}$ molecules per second, see Table~\ref{ObsTable}, using the same model as for R1 Lovejoy.

\section{Discussion}
While the chemical reactions taking place in a cometary coma can play an important role in the radial distribution of molecules \citep{RodgersCharnley05}, HCN is considered mainly as a parent species, originating from the nucleus itself.
Recent ALMA (Atacama Large Millimiter Array) spectral maps of the HCN distribution in two cometary comae are indeed consistent with centrally peaked, symmetric distributions \citep{Cordiner14}, further supporting its designation as a parent species. This, together with the observed double-peaked line profiles, to some extent validates the use of the homogeneously expanding Haser model for estimating the HCN production rates presented here.

\subsection{R1 Lovejoy}
\begin{figure}
   \centering
   \includegraphics[width=\hsize]{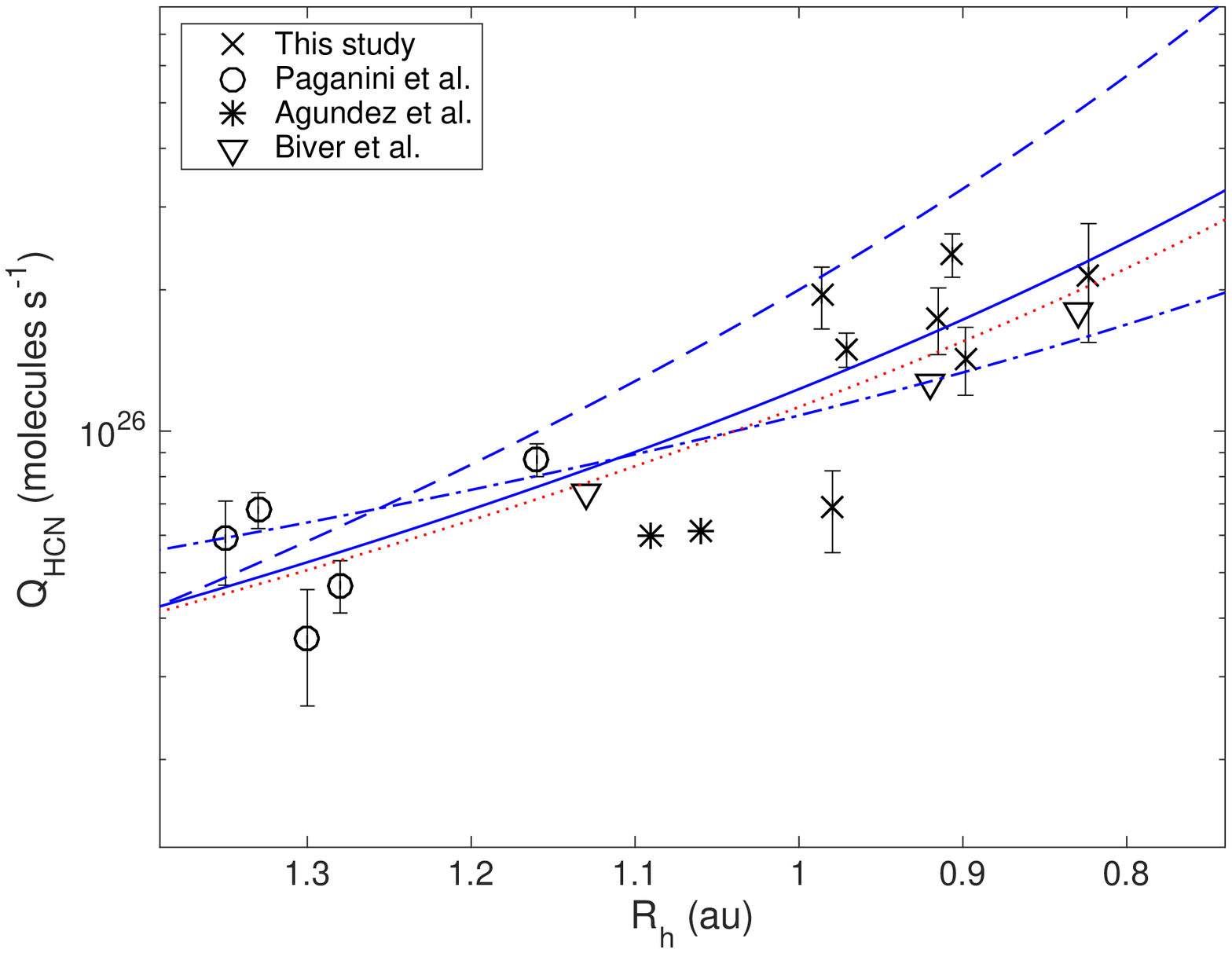}
   \caption{HCN production rates as a function of heliocentric distance for comet C/2013 R1 Lovejoy. Error bars shown for the current data (crosses) does not include uncertainties in model parameters, but only reflect measurement errors.    Circles represent data from \citet{Paganini14}, stars data from \citet{Agundez14}, and triangles data from \citet{Biver14}. The solid line illustrates the best fit to the current data, $Q_{\rm HCN} = 1.23 \times 10^{26} \times R_h^{-3.2}$~s$^{-1}$, while  the dashed line represents $Q_{\rm HCN}  \propto R_h^{-4.7}$, the dash-dotted line $Q_{\rm HCN} \propto R_h^{-2.0}$, and the dotted line $Q_{\rm HCN} \propto R_h^{-3.0}$, see text for details.}
              \label{FigProdRates}
    \end{figure}
Figure~\ref{FigProdRates} shows our estimated HCN production rates in relation to previously published ones, as a function of heliocentric distance. Our production rates are consistent with an increase in activity with decreasing heliocentric distance, best fit by the power law function $Q_{\rm HCN} = 1.23 \times 10^{26} \times R_h^{-3.2}$~s$^{-1}$, shown as a solid line in Fig.~\ref{FigProdRates}.  Compared to production rates reported by \citet{Biver14} from the same period of time (note that the errorbars of the latter dataset are too small to display at the scale of the plot), current rates are in general found to be higher. 

The production rate variability implied by the current dataset has an exponential increase, $R_h^{-3.2}$, that is steeper than what is typical for comets \citep[$R_h^{-2.0}$, ][]{Biver99}. For comparison,
the dash-dotted line in Fig.~\ref{FigProdRates} represents a production rate variation $Q_{\rm HCN} = Q_{\rm HCN}(1~{\rm au})\,R_h^{-2.0}$, with $Q_{\rm HCN}(1~{\rm au})=1.08 \times 10^{26}$~s$^{-1}$ to fit the first of the \citet{Paganini14} data points. This trend fits the \citet{Biver14} data rather well. On the other hand, based on their near-infrared data, \citet{Paganini14} reports an even steeper increase in water production rate, with $Q_{\rm H_2O} = 6.66 \times 10^{28} \times R_h^{-4.7}$~s$^{-1}$, and an HCN abundance of around 0.3\% with respect to water. The dashed line in Fig.~\ref{FigProdRates} represent such production rate variation for HCN with $Q_{\rm HCN}(1~{\rm au})=0.003 \times 6.66 \times 10^{28} = 2.00 \times 10^{26}$~s$^{-1}$. We note that all production rates reported from later dates, including those of the present study, falls below this predicted trend. In fact, when assigning all data points included in Fig.~\ref{FigProdRates} equal weight, the power law that best reproduces the HCN production rate variablility of R1 Lovejoy is $Q_{\rm HCN} = 1.13 \times 10^{26} \times R_h^{-3.0}$~s$^{-1}$ (dotted line in Fig.~\ref{FigProdRates}), very similar to that derived from the current dataset only.

Thus, even though the derived HCN production rates varies by factors of a few over timescales of days, the trend they represent follow the general variation of the comet. Whether the substantial variations is an effect of a rotating, inhomogeneous nucleus, or overall fluctuation in nucleus activity, cannot be determined. No periodicity can be discerned either in the variations presented here, or over sub-sets of data from days with longer integration times. We note that the Paganini data also show substantial variation over days, albeit within a factor of two.

This study implies a global HCN abundance relative to water in R1 Lovejoy of 
1.0--2.8$\times 10^{-3}$ with a mean of 0.2\%, somewhat lower than reported by \citet{Paganini14} (0.3\%), but within the range of what is measured for comets in general \citep[0.09--0.5\%;][]{MummaCharnley11}.

\subsection{Q2 Lovejoy}
The HCN production rate of comet Q2 Lovejoy on the single day of observations was higher than what we observe from R1 Lovejoy over three weeks approaching perihelion. More noteworthy is that, compared to the production rates of R1 Lovejoy at similar heliocentric distances around 1.3 au, determined by \citet{Paganini14}, Q2 Lovejoy was producing more than 5 times more HCN, indicating a significantly higher level of volatile outgassing from the nucleus. Corroborating this, it was concluded by \citet{Biver15} that Q2 Lovejoy was one of the most active comets in the last two decades, enabling them to detect a suite of complex organic molecules in its coma. 

Our observations imply a global HCN abundance relative to water of $9\times 10^{-4}$, the same as \citep{Biver15} reports being the average over their observing period. This demonstrates the excellent performance of the available receiver system at the OSO20m for comet observations.

\section{Conclusions}
We have presented Onsala 20 m observations of HCN J=1--0 in comets C/2013 R1 (Lovejoy) and C/2014 Q2 (Lovejoy) within a month of their respective perihelion passages. Our estimated HCN production rates for R1 Lovejoy between November 22 and December 15, at 0.7 -- $2.4\times 10^{26}$ molecules s$^{-1}$, are somewhat higher than previously reported for overlapping dates \citep{Biver14}, but the implied increase in production rate with heliocentric distance, $Q_{\rm HCN}  \propto R_h^{-3.2}$, represent well the overall increase over the time it has been observed, from $R_h$=1.35 \citep{Paganini14} and inward. The implied mean HCN abundance relative to water in R1 Lovejoy is 0.2\%.   

The HCN production rate we estimate for Q2 Lovejoy at a heliocentric distance of 1.3 au, $\sim 4.5 \times 10^{26}$ molecules s$^{-1}$, is more than 5 times higher relative to comet R1 Lovejoy at similar heliocentric distances. The implied mean HCN abundance relative to water, 0.09\%, is the same as reported by \citet{Biver15} for an observing period including the date of our observations.

The comparison in data quality between the presented cometary observations with the OSO20m demonstrates the high performance of the upgraded 3mm receiver system and its potential in future coma monitoring of relatively bright comets.

\begin{acknowledgements}
We sincerely thank Nicolas Biver for being gracious enough to assist in modeling the non-LTE radiative transfer and who's comments have greatly improved the paper. We are also grateful to Stefanie Milam for useful advice on the analysis. 
\end{acknowledgements}


\bibliographystyle{aa}
\bibliography{references}

\end{document}